\documentclass{article}
\usepackage{spconf,amsmath,graphicx}

\usepackage{microtype}
\usepackage{graphicx}
\usepackage{subfigure}
\usepackage{booktabs} 

\usepackage{enumerate}
\usepackage{subfigure}
\usepackage{cite}
\usepackage{multirow}

\usepackage{hyperref}
\usepackage[dvipsnames]{xcolor}

\usepackage{amsmath,amsfonts,bm}

\def\eqref#1{equation~\ref{#1}}

\def\1{\bm{1}}

\def\vv{{\bm{v}}}

\def\vx{{\bm{x}}}

\def\vz{{\bm{z}}}

\def\mC{{\bm{C}}}

\def\mV{{\bm{V}}}

\DeclareMathAlphabet{\mathsfit}{\encodingdefault}{\sfdefault}{m}{sl}
\SetMathAlphabet{\mathsfit}{bold}{\encodingdefault}{\sfdefault}{bx}{n}

\newcommand{\E}{\mathbb{E}}

\definecolor{amethyst}{rgb}{0.6, 0.4, 0.8}

\let\OLDthebibliography\thebibliography
\renewcommand\thebibliography[1]{
  \OLDthebibliography{#1}
  \setlength{\parskip}{1.4pt plus 0.3ex}
  \setlength{\itemsep}{1.4pt plus 0.3ex}
}

\newcommand{\model}{SELFIE}

\title{Improving Self-Supervised Learning for Audio Representations\linebreak by Feature Diversity and Decorrelation}
\name{Bac Nguyen, Stefan Uhlich, Fabien Cardinaux}
\address{Sony Europe B.V., R\&D Center, Stuttgart Laboratory 1, Germany}
\begin{document}

\ninept

\setlength{\abovedisplayskip}{2pt}
\setlength{\belowdisplayskip}{2pt}
\setlength{\abovedisplayshortskip}{2pt}
\setlength{\belowdisplayshortskip}{2pt}

\maketitle
\begin{abstract}
Self-supervised learning (SSL) has recently shown remarkable results in closing the gap between supervised and unsupervised learning. The idea is to learn robust features that are invariant to distortions of the input data. Despite its success, this idea can suffer from a collapsing issue where the network produces a constant representation. To this end, we introduce \model{}, a novel \textbf{Se}lf-supervised \textbf{L}earning approach for audio representation via \textbf{F}eature D\textbf{i}versity and D\textbf{e}correlation. \model{} avoids the collapsing issue by ensuring that the representation (i) maintains a high diversity among embeddings and (ii) decorrelates the dependencies between dimensions. \model{} is pre-trained on the large-scale AudioSet dataset and its embeddings are validated on nine audio downstream tasks, including speech, music, and sound event recognition. Experimental results show that \model{} outperforms existing SSL methods in several tasks. 
\end{abstract}

\section{Introduction}
Deep learning has achieved great success in many application domains, including computer vision, audio processing, text processing, and others~\cite{Goodfellow-et-al-2016}. However, it often requires a huge amount of labeled data for training. To reduce the expensive cost of annotating large-scale data, self-supervised learning (SSL) aims to learn powerful feature representations by leveraging the supervisory signals from the input data itself. Learning is often done by solving a hand-crafted pre-text task without using any human-annotated labels. Various pre-text tasks for SSL have been proposed, including prediction of future frames~\cite{chung2019unsupervised}, masked feature prediction~\cite{baevski2020wav2vec,Hsu2021HuBERT}, contrastive learning~\cite{saeed2020cola,wang2022universal,al2021clar,coala_2020}, and predictive coding~\cite{oord2018representation}. Once the network is trained to solve the pre-text task, feature representations are extracted from the pre-trained model in order to solve new downstream tasks. Powerful and generic representations can benefit downstream tasks, especially those with limited labeled data.

A general framework for SSL is based on maximizing the correlations between different views of the same object in a latent space~\cite{liu2022audio}. By learning invariant features to distortions of the input, an SSL model is expected to extract high-level semantic information. Depending on the use of negative examples (\textit{i.e.,} views from other objects), SSL methods can be categorized into two groups: contrastive and predictive learning. In contrastive learning, embeddings of positive examples are mapped close together, while those of negative examples are mapped far apart. Positive pairs can be audio segments from the same clip and negative pairs can be audio segments from different clips~\cite{saeed2020cola,fonseca2020uclser20}. Negative examples are used to encourage the diversity of representations, preventing the trivial constant embedding. On the other hand, predictive models only maximize the similarity between positive examples without considering any negative examples. To avoid model collapse, several techniques have been proposed. For instance, BYOL~\cite{grill2020byol} predicts the output of one view from another view and employs a momentum encoder to maintain consistent representations. SimSiam~\cite{chen2021exploring} introduces a stop-gradient operation on the target branch. VICReg~\cite{bardes2021vicreg} maintains a large variance for each embedding dimension within a batch for diversity.

Learning representations for audio and speech has recently gained popularity due to its promising performance over diverse challenging tasks. For instance, HuBERT~\cite{Hsu2021HuBERT} and wav2vec 2.0~\cite{baevski2020wav2vec} show state-of-the-art results for automatic speech recognition (ASR) with limited training data. Other SSL methods obtain significant improvements on non-sematic speech tasks, including TRILL~\cite{shor2020trill}, COLA~\cite{saeed2020cola}, Conformer-based models~\cite{shor2022universal}, and BYOL-A~\cite{niizumi2022byol} . 
Among the most successful SSL methods, contrastive learning has been widely applied for speech and sound events~\cite{fonseca2020uclser20,saeed2020cola}. For instance, \cite{jansen2017triplet} adopted the triplet loss in metric learning to learn audio representations. Triplets were constructed by simply applying audio augmentations. Alternatively, CPC~\cite{oord2018representation} predicted the future information based on the global context from the past using an auto-regressive model.  Despite its popularity, contrastive learning methods tend to be computationally expensive and require a large mini-batch size to converge. They also suffer from dimensional collapse, where embeddings only lie on a lower-dimensional subspace instead of the entire embedding space~\cite{jing2021understanding}. There have been a few attempts that do not use contrastive learning. HuBERT~\cite{Hsu2021HuBERT} relied on BERT-like pre-training with the pseudo-labels provided by $k$-means clustering. BYOL-A~\cite{niizumi2022byol} was adopted from BYOL~\cite{grill2020byol}, which learned representations without using negative samples. Although empirical studies showed promising results, how these methods can avoid collapsing issue was not fully addressed.

The key requirements for extracting good representations include three aspects. (R1) First, the learned representation should be invariant to distortions of the input created by data augmentations. This encourages the representation to capture high-level semantic information. (R2) Second, it should  facilitate a learning system on a downstream task by extracting useful information from data. (R3) Third, it should not contain redundant information (informational collapse)~\cite{bardes2021vicreg,reducing_overfitting} in which the dimensions of the embeddings are highly correlated. The last requirement encourages disentangled representations. Following these guiding principles, we propose a novel SSL approach to learn good feature representations.



In summary, the contributions of our work are as follows. (i) We introduce \model{}, a simple approach to learn general-purpose representations of sounds. Our method maximizes the similarity between two augmented views of audio mel-spectrograms. Several data augmentation techniques are employed to create different views. To avoid the collapsed representations, we force the embeddings within the same mini-batch to be different by maximizing the distances between them. In addition, we prevent informational collapse be decorrelating all pairs of dimensions. (ii) We demonstrate the performance of our approach over nine downstream audio tasks using linear evaluation. More specifically, a linear classifier is trained on top of the pre-trained embeddings to evaluate a downstream task. The results show that \model{} is able to learn useful audio representations.

\section{Proposed method}
In this section, we describe our proposed SSL method and its training objective functions, followed by implementation details.

\begin{figure*}[t]
    \centering
    \vspace{-0.2cm}
    \includegraphics[width=0.95\textwidth]{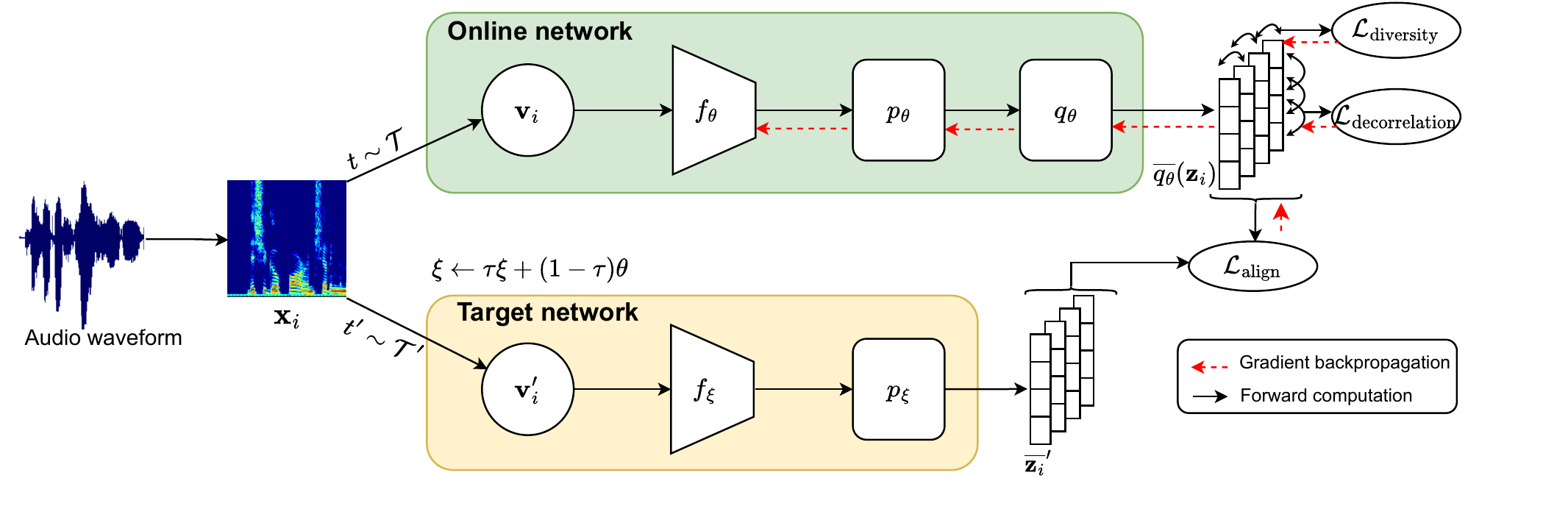}
    \vspace{-0.3cm}
    \caption{\model{}: Self-supervised learning via feature diversity and decorrelation. It consists of an online network and a target network. Both networks share the same architecture, but use different weights. The online network is trained by stochastic gradient descent, while the target network is updated using a slowly moving exponential average.}
    \label{fig:selfd}
    \vspace{-0.2cm}
\end{figure*}

\subsection{\model{}}
Let  $\vv_i = t(\vx_i)$ and $\vv_i' = t'(\vx_i)$ be two views from different random data augmentations of an input $\vx_i$, where $t$ and $t'$ are transformations sampled from two data augmentation distributions $t\sim \mathcal{T}$ and $t' \sim \mathcal{T}'$. \model{} shares the same training scheme as the recent BYOL method~\cite{grill2020byol}. Fig.~\ref{fig:selfd} illustrates the overall network architecture and the loss functions of \model{}. First, two views $\vv_i$ and $\vv_i'$ are encoded by an online network $f_{\theta}$ and a target network $f_{\xi}$, respectively. Subsequently, they are fed to projector networks $p_{\theta}$ and $p_{\xi}$, which output projections $\vz_i = p_{\theta}(f_{\theta}(\vv))$ and $\vz_i' = p_{\xi} (f_{\xi} (\vv'))$. Unlike the target network, the online network has in addition a predictor network $q_{\theta}$, which then produces a prediction $q_{\theta}(\vz_i)$. Finally, we normalize the outputs from the online network as well as the target network, \textit{i.e.,} $\overline{\vz_i'} = \vz_i'/\| \vz'\|_2$ and $\overline{q_{\theta}} (\vz_i)= q_{\theta} (\vz_i) / \| q_{\theta} (\vz_i)\|_2$. Note that both networks share the same architecture but use different weights. In particular, the online network is trained using gradient descent, while the target network is updated using a slowly moving exponential  average,
\begin{align*}
    \xi \leftarrow \tau\xi + (1-\tau) \theta\,,
\end{align*}
where $\tau \in [0,1]$ is the chosen decay rate. Next, we describe the training objectives, which are used to train the online network. 

\textbf{Alignment loss.} Our first requirement (R1) is to ensure the invariance to data augmentations.  This requirement can be satisfied by minimizing the alignment loss, which is defined as the squared Euclidean distance between the normalized predictions and target projections,
\begin{align}
     \mathcal{L}_{\text{align}} = \E_{\vz_i, \vz_i'} \Big[ \big\| \overline{q_{\theta}} (\vz_i) - \overline{\vz_i'}\big\|_2^2\Big] \,. \label{eq:aligment}
\end{align}
By minimizing the alignment loss, the learned representation should extract the most relevant information shared between positive examples and remain invariant to noises. As empirically shown by Grill et al.~\cite{grill2020byol}, the predictor $q_{\theta}$ and the momentum encoder can help to alleviate the collapsing issue as well as to stabilize the training, specially when the mini-batch size is small.

\textbf{Diversity loss.} Our second requirement (R2) is to ensure that the learned representation provides useful information for downstream tasks. We introduce the diversity loss $\mathcal{L}_{\text{diversity}}$, which maximizes the distances between all normalized predictions of the online network,
\begin{align}
    \mathcal{L}_{\text{diversity}} = \E_{\vz_i, \vz_j}\Big[ -\big\| \overline{q_{\theta}} (\vz_i) - \overline{q_{\theta}}(\vz_j) \big\|^2_2 \Big] \,.\label{eq:diversity}
\end{align}
By increasing the diversity in embeddings, it helps to capture different structural information of data, which might be useful for different downstream tasks. Another benefit is that the diversity loss prevents embeddings from collapsing to the same representation. Our diversity loss is closely related to the notion of uniformity introduced by Wang and Isola~\cite{wang2020understanding}. A theoretical justification of diversity has been given in \cite{laakom2021feature}. Let $n$ denote the mini-batch size and $\mV \in \mathbb{R}^{d\times n}$ denote the matrix containing the $n$ normalized prediction vectors, where each has a dimension of $d$. A naive implementation of Eq.~(\ref{eq:diversity}) has a complexity of $\mathcal{O}(n^2d)$ as it requires the computation of distances between all possible pairs. Because the projections are normalized, we can simplify  $\mathcal{L}_{\text{diversity}}$ as follows
\begin{align*}
   \mathcal{L}_{\text{diversity}} &= \frac{1}{n^2}\Big[\mathbf{e}^{1\times n}(\mathbf{e}^{n\times n} - 2\mV^\top\mV + \mathbf{e}^{n\times n})\mathbf{e}^{n\times 1}\Big] \\
   &= 2 - \frac{2}{n^2}(\mV\mathbf{e}^{n\times 1})^\top(\mV\mathbf{e}^{n\times 1}) =  2 - \frac{2}{n^2}\big\|\mV\mathbf{e}^{n\times 1}\big\|_2^2\,, 
\end{align*}
where $\textbf{e}^{n\times m} \in \mathbb{R}^{n\times m}$ is a matrix whose entries are all 1's. Therefore, we can reduce the computational complexity to $\mathcal{O}(nd)$.

\textbf{Decorrelation loss.} Our third requirement (R3) is to ensure that  each feature dimension encodes different information to avoid the informational collapse issue. Inspired by VICReg~\cite{bardes2021vicreg}, we use the decorrelation loss defined as
\begin{align}
    \mathcal{L}_{\text{decorrelation}} = \sum_{i \ne j} \Big[\mC\big(\overline{q_{\theta}}(\vz)\big)_{ij}\Big]^2 \,, \label{eq:decorrelation}
\end{align}
where $\mC(\overline{q_{\theta}}(\vz))$ is the covariance matrix of normalized predictions
\begin{align*}
   \mC\big(\overline{q_{\theta}}(\vz)\big) = \frac{1}{n - 1} \sum_{\vz_i} \big(\overline{q_{\theta}}(\vz_i) - \widehat{q_{\theta}}(\vz)\big)\big(\overline{q_{\theta}}(\vz_i) - \widehat{q_{\theta}}(\vz)\big)^\top
\end{align*}
and $\widehat{q_{\theta}}(\vz)$ denotes the mean vector
\begin{align*}
   \widehat{q_{\theta}}(\vz) = \frac{1}{n} \sum_{\vz_i} \overline{q_{\theta}}(\vz_i) \,. 
\end{align*}
The decorrelation loss aims to minimize the off-diagonal coefficients of $\mC(\overline{q_{\theta}}(\vz))$, which prevent feature dimensions to encode similar information.

Using Eqs.~(\ref{eq:aligment}) to (\ref{eq:decorrelation}), our final loss function is defined as
\begin{align*}
    \mathcal{L} = \mathcal{L}_{\text{align}} +  \lambda_{\text{diversity}}\mathcal{L}_{\text{diversity}} + \lambda_{\text{decorrelation}}\mathcal{L}_{\text{decorrelation}}\,,
\end{align*}
where $\lambda_{\text{diversity}} \ge 0$ and $\lambda_{\text{decorrelation}} \ge 0$ are weighing terms. In our experiments, these terms are set to $\lambda_{\text{diversity}}=1$ and $\lambda_{\text{decorrelation}}=1$.

\subsection{Implementation details}
Although raw audio waveforms can be input to the encoder network, we use time-frequency features to learn representations due to their great improvements over raw audio signals~\cite{al2021clar,7952651}.    Given an audio clip, we compute the log mel-spectrogram with a window size of 64 ms, a hop size of 10 ms, and 64 mel-filterbanks. These log mel-spectrograms are considered as inputs to the network.  In the following, we describe in detail the network architecture of the encoder and data augmentation strategies used in our implementation.

\textbf{Network architecture.} We use a simple and lightweight encoder network proposed in BYOL-A~\cite{niizumi2022byol} to extract feature representations. The network architecture is based on the audio embedding block~\cite{koizumi2020t6ntt}, which consists of two convolutional neural network (CNN) blocks and two fully-connected layer (FC) blocks. A kernel size of $3\times 3$ is used for all CNN layers, followed by a batch normalization layer and a max pooling layer of kernel size $2\times 2$ with stride of 2. We use ReLU as the activation function. The outputs from the CNN blocks are considered as local features. These outputs are reshaped to flatten the frequency and channel dimensions, which will be fed to the FC blocks to learn global features. We then concatenate the outputs from the CNN blocks as well as the FC blocks to combine both local and global features. Finally, a mean-max pooling operation is applied to get an embedding of dimension 3072. The projector and predictor networks are MLPs with a single hidden layer of size 4096, followed by a batch normalization layer, ReLU as the activation function, and a linear layer to output an embedding of dimension 256. The decay rate for the exponential moving average is set to $\tau = 0.995$.  In our implementation, we replace the Pre-Norm block in~\cite{niizumi2022byol} by a batch normalization layer and do not use any Post-Norm.

\textbf{Data augmentations.} We use three strategies for data augmentations. First, random resize crop (RRC) is applied to the mel-spectrogram input with a random cropped area of scale between 0.6 and 1.5 and an aspect ratio of the crop between 0.75 and 1.33. Although RRC is mainly used for computer vision, it can be considered as an approximation for pitch/time shift and stretch. In other words, representations that are invariant to RRC will be invariant to perturbations of pitch and time~\cite{niizumi2022byol}. Second, we apply random background noise (RBN) by mixing random pairs of inputs~\cite{zhang2018mixup}. Given a mel-spectrogram $\vx$, we compute a linear combination with another input $\vx'$ in the same mini-batch with a small mixing ratio $\lambda \sim \mathcal{U}(0, 0.2)$, \textit{i.e.,} $(1 - \lambda)\vx + \lambda \vx'$. RBN enables the representation to be more robust to noise. Finally, random linear fader~\cite{niizumi2022byol,al2021clar} (RLF) applies a linear change in volume to the audio input, which simulates fade-in/out effects.

\section{Experiments}
In this section, we validate the effectiveness of \model{} on various diverse downstream tasks. First, \model{} is pre-trained on a large-scale audio dataset, then its embeddings are transferred to downstream tasks. A linear classifier is trained on top of the frozen embeddings. Classification accuracy is reported as the performance for each task.

\subsection{Datasets and downstream tasks}
We train \model{} on the popular AudioSet~\cite{gemmeke2017audioset} dataset. It consists of two million audio clips extracted from YouTube videos. The dataset has very diverse categories, covering a large range of human, animal, music, and environmental sounds. Although AudioSet provides multi-label annotations for each audio clip, this information is not used for our training. All sound clips are sampled at a rate of 16 kHz. Inputs to \model{} are clips of length 0.95 seconds, which are randomly cropped from the audio input. 

Table~\ref{tab:downstreamtask} summarizes the downstream tasks used in our experiments. These datasets are commonly reported by previous studies for evaluating SSL methods in audio domain~\cite{saeed2020cola,niizumi2022byol}. The benchmark includes two sound event recognition (SER) tasks, four non-semantic speech (NOSS) tasks, and three music tasks. We briefly describe them as follows. ESC-50~\cite{piczak2015esc50} is a sound classification problem, containing 50 environmental sound classes.  UrbanSound8K~\cite{salamon2014urbansound} (US8K) is another sound classification problem, containing ten urban sound classes. Leave-one-out cross-validation is employed with the official folds to report the accuracy for these two tasks. Speech Commands V2~\cite{speechcommandsv2} (SPCV2) is a word classification task, containing 35 spoken commands (classes) from one second of audio. VoxCeleb1~\cite{voxceleb} (VC1) is a speaker identification task, containing 1,251 speakers extracted from interview videos of celebrities. VoxForge~\cite{voxforge} (VF) is the task of identifying the language being spoken from an audio input. Spoken audio are from six languages (English, French, German, Spanish, Russian, and Italian). CREMA-D~\cite{cao2014cremad} (CRM-D) is a speech emotion recognition task, containing 6 classes (anger, disgust, fear, happy, neutral, and sad).  GTZAN~\cite{gt2013gtzan} is a music genre recognition task, containing 10 music genre classes. NSynth~\cite{nsynth2017} is an instrument family classification task, containing 11 classes. Surge~\cite{turian2021torchsynth} is a pitch audio classification task, containing 88 MIDI note classes. We follow the same protocol as in~\cite{niizumi2022byol} for training, validation and testing. Audio inputs in downstream tasks are randomly cropped to the average duration length for linear evaluation. All results are averaged over five runs using different random seeds for training \model{}. 

\begin{table}[t]
    \centering
    \vspace{-0.5cm}
    \caption{Description of datasets used for downstream tasks.}
    \label{tab:downstreamtask}
    \resizebox{\linewidth}{!}{
    \begin{tabular}{lccrr}
    \toprule
    Dataset     & Task & \# of classes & \# of examples & Avg length  \\
    \midrule
     ESC-50~\cite{piczak2015esc50}    & SER & 50 & 2,000 & 5.0 s\\
     US8K~\cite{salamon2014urbansound}    & SER & 10 & 8,732 & 4.0 s\\
     SPCV2~\cite{speechcommandsv2}    & NOSS & 35 & 105,829& 1.0 s\\
     VC1~\cite{voxceleb}    & NOSS & 1,251& 153,516 & 8.2 s\\
     VF~\cite{voxforge}    & NOSS & 6 & 176,428 & 5.8 s\\
     CRM-D~\cite{cao2014cremad}    & NOSS & 6 & 7,438 & 2.5 s\\
     GTZAN~\cite{gt2013gtzan}    & Music & 10 & 930 & 30.0 s\\
     NSynth~\cite{nsynth2017}    & Music & 11 & 305,979& 4.0 s\\
     Surge~\cite{turian2021torchsynth}    & Music & 88 & 183,392 & 4.0 s\\
    \bottomrule
    \end{tabular}
    }
    \vspace{-0.5cm}
\end{table}

\subsection{Results on dowstream tasks}

\begin{table*}[htbp]
\caption{Linear evaluation accuracies (\%) with 95\% confidence intervals (CI) on downstream tasks. \model{} is pre-trained on AudioSet~\cite{gemmeke2017audioset}.}
\label{tab:result-benchmark}
\centering
\resizebox{\textwidth}{!}{%
\begin{tabular}{lllllllllll}\toprule
\multirow{2}{*}{Method}&  \multicolumn{2}{c}{SER tasks} & \multicolumn{4}{c}{NOSS tasks} & \multicolumn{3}{c}{Music tasks} & \\
\cmidrule(lr){2-3} \cmidrule(lr){4-7} \cmidrule(lr){8-10}  
 &    ESC-50 &    US8K &    SPCV2 &    VC1 &     VF &    CRM-D &    GTZAN &     NSynth &      Surge & Average \\
\midrule
COLA{\cite{saeed2020cola}}   &   N/A &   N/A &  62.4 &  29.9 &  71.3 &   N/A &   N/A &  63.4 &   N/A &   N/A \\
mel-spectrogram   &  16.9 {\fontsize{6pt}{6pt}\selectfont $\pm$ 1.9} &   43.7{\fontsize{6pt}{6pt}\selectfont $\pm$ 0.3} &  24.8 {\fontsize{6pt}{6pt}\selectfont $\pm$ 0.1} &  9.2 {\fontsize{6pt}{6pt}\selectfont $\pm$ 0.1} &  66.1 {\fontsize{6pt}{6pt}\selectfont $\pm$ 0.1} &  40.4 {\fontsize{6pt}{6pt}\selectfont $\pm$ 1.0} &   32.4 {\fontsize{6pt}{6pt}\selectfont $\pm$ 4.1} &  33.4 {\fontsize{6pt}{6pt}\selectfont $\pm$ 0.8} &   32.9 {\fontsize{6pt}{6pt}\selectfont $\pm$ 0.1} &   33.2 \\
\midrule
TRILL\cite{shor2020trill}       &  75.4 {\fontsize{6pt}{6pt}\selectfont $\pm$ 0.7} &  75.2 {\fontsize{6pt}{6pt}\selectfont $\pm$ 1.3} &  78.4 {\fontsize{6pt}{6pt}\selectfont $\pm$ 0.8} &  40.1 {\fontsize{6pt}{6pt}\selectfont $\pm$ 1.1} &  88.8 {\fontsize{6pt}{6pt}\selectfont $\pm$ 0.3} &  58.8 {\fontsize{6pt}{6pt}\selectfont $\pm$ 2.3} &  64.4 {\fontsize{6pt}{6pt}\selectfont $\pm$ 1.8} &  \textbf{74.3 {\fontsize{6pt}{6pt}\selectfont $\pm$ 1.8}} &  28.7 {\fontsize{6pt}{6pt}\selectfont $\pm$ 1.0} & 64.9\\
Wav2Vec2-F\cite{baevski2020wav2vec}  &  65.6 {\fontsize{6pt}{6pt}\selectfont $\pm$ 1.7} &  67.8 {\fontsize{6pt}{6pt}\selectfont $\pm$ 0.3} &  85.8 {\fontsize{6pt}{6pt}\selectfont $\pm$ 0.2} &  32.0 {\fontsize{6pt}{6pt}\selectfont $\pm$ 0.3} &  81.7 {\fontsize{6pt}{6pt}\selectfont $\pm$ 0.1} &  56.4 {\fontsize{6pt}{6pt}\selectfont $\pm$ 0.5} &  62.3 {\fontsize{6pt}{6pt}\selectfont $\pm$ 1.0} &  62.2 {\fontsize{6pt}{6pt}\selectfont $\pm$ 0.8} &  30.0 {\fontsize{6pt}{6pt}\selectfont $\pm$ 0.4} & 60.4\\
Wav2Vec2-C\cite{baevski2020wav2vec}  &  59.3 {\fontsize{6pt}{6pt}\selectfont $\pm$ 0.4} &  64.7 {\fontsize{6pt}{6pt}\selectfont $\pm$ 0.6} & \textbf{96.7 {\fontsize{6pt}{6pt}\selectfont $\pm$ 0.1}}&  14.3 {\fontsize{6pt}{6pt}\selectfont $\pm$ 0.2} &\textbf{98.5 {\fontsize{6pt}{6pt}\selectfont $\pm$ 0.0}}&59.4 {\fontsize{6pt}{6pt}\selectfont $\pm$ 1.3}&  54.7 {\fontsize{6pt}{6pt}\selectfont $\pm$ 1.3} &  56.3 {\fontsize{6pt}{6pt}\selectfont $\pm$ 1.2} &  13.4 {\fontsize{6pt}{6pt}\selectfont $\pm$ 0.1} & 57.5\\
BYOL-A~\cite{niizumi2022byol}      &  \textbf{83.2 {\fontsize{6pt}{6pt}\selectfont $\pm$ 0.6}} &  \textbf{79.7 {\fontsize{6pt}{6pt}\selectfont $\pm$ 0.5}} &  93.1 {\fontsize{6pt}{6pt}\selectfont $\pm$ 0.4} &\textbf{57.6 {\fontsize{6pt}{6pt}\selectfont $\pm$ 0.2}}&  93.3 {\fontsize{6pt}{6pt}\selectfont $\pm$ 0.3} &  63.8 {\fontsize{6pt}{6pt}\selectfont $\pm$ 1.0} &  70.1 {\fontsize{6pt}{6pt}\selectfont $\pm$ 3.6} &  73.1 {\fontsize{6pt}{6pt}\selectfont $\pm$ 0.8} &  \textbf{37.6 {\fontsize{6pt}{6pt}\selectfont $\pm$ 0.3}} & 72.4\\
\midrule
\model{} & 82.5 {\fontsize{6pt}{6pt}\selectfont $\pm$ 0.7} & 78.7 {\fontsize{6pt}{6pt}\selectfont $\pm$ 0.2} & 94.0{\fontsize{6pt}{6pt}\selectfont $\pm$ 0.1} &  55.9{\fontsize{6pt}{6pt}\selectfont $\pm$ 0.2} & 93.8 {\fontsize{6pt}{6pt}\selectfont $\pm$ 0.2} & \textbf{65.4 {\fontsize{6pt}{6pt}\selectfont $\pm$ 0.8}} & \textbf{73.6 {\fontsize{6pt}{6pt}\selectfont $\pm$ 1.1}} & 74.0 {\fontsize{6pt}{6pt}\selectfont $\pm$ 0.5} & 36.4 {\fontsize{6pt}{6pt}\selectfont $\pm$ 0.1} & \textbf{72.7}  \\ 
\bottomrule
\end{tabular}
}
\vspace{-0.5cm}
\end{table*}

\begin{table*}[htbp]
\centering
\caption{Linear evaluation accuracies (\%) on downstream tasks. All models are pre-trained on  FSD50K~\cite{fonseca2020fsd50k}.}
\label{tab:ablation_losses}
\resizebox{0.85\textwidth}{!}{
\begin{tabular}{lllllllllll}
\toprule
\multirow{2}{*}{Method} &   \multicolumn{2}{c}{SER tasks} & \multicolumn{4}{c}{NOSS tasks} & \multicolumn{3}{c}{Music tasks} & \multirow{2}{*}{Average}\\
\cmidrule(lr){2-3} \cmidrule(lr){4-7} \cmidrule(lr){8-10}  
 &    ESC-50 &    US8K &    SPCV2 &    VC1 &     VF &    CRM-D &    GTZAN &     NSynth &      Surge & \\
\midrule
BYOL-A~\cite{niizumi2022byol} &  82.5 &  \textbf{78.8} &  91.5 &51.4&  91.4 &  58.5 &  65.1 & \textbf{75.5}&  \textbf{38.3} & 70.3\\
\midrule
\model{} & \textbf{83.1} & 78.0 & \textbf{92.7}& \textbf{54.1}& \textbf{92.5} &\textbf{63.9} & \textbf{69.2}& 74.4& 36.6 & \textbf{71.6} \\
\midrule
$\mathcal{L}_{\text{align}}$ and $\mathcal{L}_{\text{diversity}}$ & 82.4 & 78.3 & 92.5 & 51.1 & 92.2 & 63.0 & 64.8& 75.5& 37.0 & 70.8 \\
$\mathcal{L}_{\text{align}}$ and $\mathcal{L}_{\text{decorrelation}}$ & 82.9 & 78.1 & 92.9 & 53.2& 92.6 & 62.7 & 66.2& 75.0 & 36.8 & 71.1 \\
$\mathcal{L}_{\text{align}}$ & 80.4 & 77.4 & 91.8& 50.1& 91.5 & 59.3 & 69.8& 75.6& 38.3 & 70.5 \\
\bottomrule
\end{tabular}
}
\vspace{-0.2cm}
\end{table*}

We compare \model{} with several baseline audio SSL methods, including TRILL~\cite{shor2020trill}, COLA~\cite{saeed2020cola}, Wav2Vec2-C~\cite{baevski2020wav2vec}, Wav2Vec2-F~\cite{baevski2020wav2vec}, and BYOL-A~\cite{niizumi2022byol}. Additionally, we also report the performances using log mel-spectrogram features. Wav2Vec2-C and Wav2Vec2-F are the 512-d embeddings from the CNN encoder and the 1,024-d embeddings from the Transformer network of wav2vec 2.0~\cite{baevski2020wav2vec}. We use the weights of wav2vec 2.0 released by Hugging Face\footnote{https://huggingface.co/facebook/wav2vec2-large-960h-lv60}, which are pre-trained on the LibriSpeech~\cite{Panayotov2015LibrispeechAA} corpus. Except Wav2Vec2-C and Wav2Vec2-F, other competing methods are pre-trained on AudioSet. \model{} is trained for 100 epochs using the Adam optimizer with a learning rate of 0.0001 and a mini-batch size of 512.

The performances over nine downstream tasks are reported in Table~\ref{tab:result-benchmark}. Overall, all SSL methods outperform the baseline using mel-spectrogram features. \model{} achieves the highest average accuracy of 72.7~\% over all tasks.  In particular, it obtains the best accuracy on the CRM-D and GTZAN datasets with a large margin.  \model{} performs competitively with its BYOL-A counterpart on the SER tasks. Wav2Vec2-F and Wav2Vec2-C show the worse results on these tasks. Since they are pre-trained only on speech data, the SER downstream tasks can be very challenging for them. On the other hand, they perform quite well on spoken language tasks such as SPCV2 and VF. Interestingly, contrastive methods like COLA and TRILL do not perform well compared to BYOL-A and \model{}.

\subsection{Ablation studies}
We empirically perform ablation studies for our method. \model{} is pre-trained on  FSD50K~\cite{fonseca2020fsd50k}, a dataset of sound events. It covers a variety of sound events from 200 classes.

\textbf{Importance of losses.} We show the impact of adding different losses to our SSL framework. The results are reported in Table~\ref{tab:ablation_losses}. We consider the result of BYOL-A reported in~\cite{niizumi2022byol} as reference. BYOL-A is equivalent to \model{} when only $\mathcal{L}_{\text{align}}$ is used (see the last row of  Table~\ref{tab:ablation_losses}). We highlight the best results (in bold) between BYOL-A and \model{}. As seen from the table, adding $\mathcal{L}_{\text{decorrelation}}$  and $\mathcal{L}_{\text{diversity}}$ improves the performance of BYOL-A. We observe that for the music tasks, it is better to use only the alignment loss. However, adding either the diversity loss or the decorelation loss can lead to great improvements for the other tasks.

\begin{figure}[t]
    \vspace{-0.4cm}
    \centering
    \includegraphics[width=0.8\linewidth]{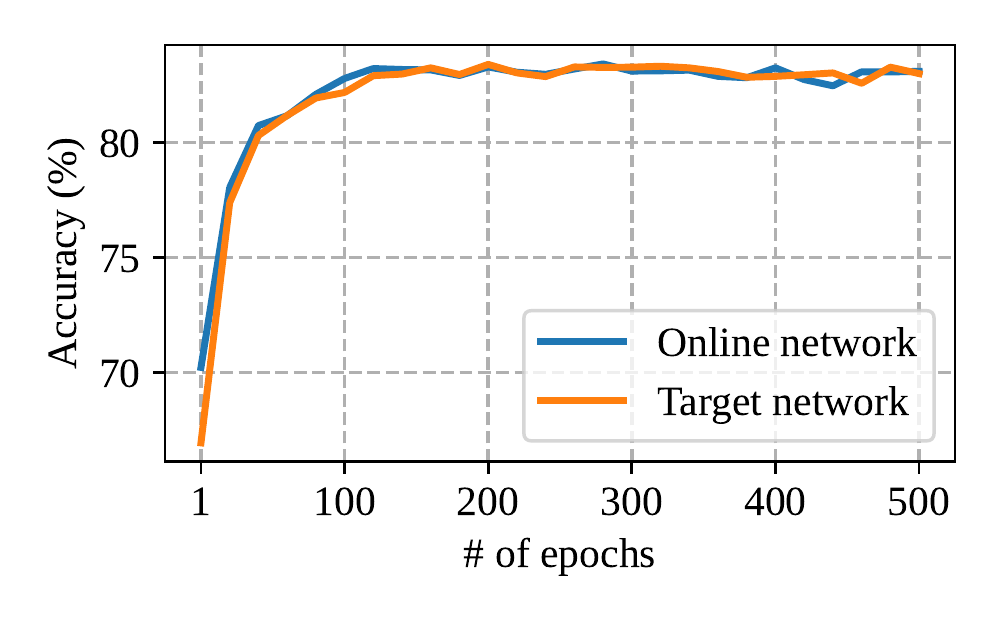}
    \vspace{-0.6cm}
    \caption{Classification accuracies over different number of epochs for the online and target networks on the ESC-50~\cite{piczak2015esc50} dataset.}
    \label{fig:esc-50}
    \vspace{-0.5cm}
\end{figure}

\textbf{Online vs target networks.} After \model{} is pre-trained, one can use embeddings computed from the target network instead of the online network for downstream tasks. For illustrative purpose, Fig.~\ref{fig:esc-50} depicts the performance of the online versus the target network on the ESC-50 dataset. They both perform quite similar. This result is expected as the predictions from the online network slowly converge to the outputs from the target network during training.

\textbf{Effects of mini-batch sizes.} We run \model{} over different mini-batch sizes, ranging from 256 to 2048. Table~\ref{tab:batch_size} reports the performance when varying mini-batch sizes. We observe that \model{} shows very stable (almost constant) performance.

\begin{table}[t]
    \centering
    \vspace{-0.2cm}
    \caption{Average accuracy (\%) of \model{} over different mini-batch sizes.}
    \label{tab:batch_size}
    \begin{tabular}{lc}
    \toprule
    Mini-batch size     &  Avg accuracy \\
    \midrule
    256 & 71.6 \\
    512 & 71.6\\
    1,024 & 71.7 \\
    2,048 & 70.6\\
    \bottomrule
    \end{tabular}
\end{table}

\section{Conclusions}
In this paper, we have introduced \model{}, a simple SSL method for audio representations, based on maximizing the similarity between different views of the same log-mel spectrogram. To avoid the collapse issue, the learned representations are forced to be diverse and contain no redundant information. Our extensive experiments over nine benchmark datasets demonstrate that \model{} is able to learn powerful audio representations.  Future work includes exploring more data augmentation techniques to further improve \model{}.

\bibliographystyle{IEEEbib}
{\small \bibliography{mybib}}
\end{document}